\documentclass[10pt,onecolumn]{article}

\pdfoutput=1

\usepackage{times}
\usepackage{graphicx}
\usepackage{amsmath}

\usepackage[text={7in,9in},centering,letterpaper]{geometry}

\usepackage{amsmath,amssymb,amsfonts}
\usepackage{algorithmic}
\usepackage{textcomp}
\usepackage{array}
\usepackage{multirow}
\usepackage{url}

\begin{document}

\title{Android Malware Detection using Feature Ranking of Permissions}

\author{Muhammad Suleman Saleem,
        Jelena Mi\v{s}i\'{c}, and
        Vojislav B. Mi\v{s}i\'{c} 
\thanks{The authors are with Ryerson University, Toronto, ON, Canada
M5B 2K3.
e-mail: \{m5saleem, jmisic, vmisic\}@ryerson.ca.}}

\maketitle
\thispagestyle{empty}

\begin{abstract}
We investigate the use of Android permissions as the vehicle to allow
for quick and effective differentiation between benign and malware
apps.  To this end, we extract all Android permissions, eliminating
those that have zero impact, and apply two feature ranking algorithms
namely Chi-Square test and Fisher's Exact test to rank and
additionally filter them, resulting in a comparatively small set of
relevant permissions. Then we use Decision Tree, Support Vector
Machine, and Random Forest Classifier algorithms to detect malware
apps. Our analysis indicates that this approach can result in better
accuracy and F-score value than other reported approaches. In
particular, when random forest is used as the classifier with the
combination of Fisher's Exact test, we achieve 99.34\% in accuracy and
92.17\% in F-score with the false positive rate of 0.56\% for the
dataset in question, with results improving to 99.82\% in accuracy and
95.28\% in F-score with the false positive rate as low as 0.05\% when
only malware from three most popular malware families are considered.
\end{abstract}

\maketitle

\hrule

\section{Introduction}
\label{sec:introduction}

In our increasingly connected society, number of mobile
devices and mobile apps continue to increase, providing numerous
services for personal and business use. Because of huge market share
and open source features, the Android platform is the main target of
hackers.  Report \cite{kelly2014report} reveals that 97\% of mobile
malware are on Android and about 8.5 million new Android malware
samples have been discovered in the first quarter of 2018 which
represents a 12\% increase from the first quarter of 2017
\cite{gdata}.

However, malware detection is a challenging task as mobile devices are
more vulnerable to malware attacks due to limited computational
resources. Antivirus solutions are generally used in desktop computers
where they track the application data in real time using the latest
malware signature files downloaded from antivirus databases, but this
is not feasible in mobile devices and does not scale well because of
significant computational overhead \cite{arp2014drebin}. 

Many approaches have been used to detect malware but they are not
always effective.  For example, DroidScribe \cite{dash2016droidscribe}
is a dynamic malware detection approach based on feature sets that
include IP addresses and ports used, network traffic size, file types,
method names, and the like. They classify malware using inter-process
communication with the combination of support vector machine-based
classifier. DroidScope \cite{yan2012droidscope} is another dynamic
malware detection approach that monitors the behavior of apps at run
time and reconstructs OS-level semantics and Java-level semantics. 

In this work, we focus on using permissions which are declared in
Android manifest file and can be extracted easily. This manifest file
contains the structure and meta data of Android app. In addition to
running each app in separate application sandbox which isolates app's
data from other apps, Android uses its permission system to enforce
additional limitations on their apps. An app can perform a specific
operation only if it has requested and subsequently obtained the
appropriate permission. Android permissions are normally grouped into
three categories, namely normal permissions, dangerous permissions and
signature permissions \cite{android-permissions}. 

\begin{itemize}

\item Normal permissions are those that pose little or no risk to the
user's privacy; they are automatically granted by the operating
system. 

\item Dangerous permissions pose higher risk to the user's privacy and
these permissions must be explicitly granted by the user. 

\item Signature permissions are granted by the system and can be
accessed by those apps only which have the same certificate as the
apps that define the permission.  Most often, signature permissions are
used by apps installed by the mobile operator.

\end{itemize}

To perform specific tasks on Android devices such as accessing WiFi,
sending SMS, reading SMS, accessing camera etc., each app has to
explicitly ask the user to grant the permissions during installation
or these permissions may be requested before using the app. End user
grant the permission and continue with the installation or deny the
permission and cancel the installation process. Effectiveness of
Android permission system is primarily dependent upon the end user who
is responsible for granting the permissions to the apps, which is
likely the weakest link in Android security mechanism. This is the
main reason why efficient and effective malware detection for Android
apps is a crucial component of the Android ecosystem.

Permissions extracted from manifest file can be used as the basis for
cost-effective malware detection \cite{wijesekera2015android} and
avoid high cost of time and computation. Our dataset is based on
Drebin dataset \cite{arp2014drebin} which contains data on nearly
120,000 app samples, including both benign and malware apps.  However,
a large number of available features makes straightforward application
infeasible, and an effective approach is necessary to reduce the
number of features.

To this end, we apply two feature ranking algorithms (Chi-Square test
and Fisher's Exact test) and automatically extract effective
permissions which have strong association with the class variable and
rank the permissions based on their p-values. Then, we apply three
machine learning algorithms, namely decision tree, random forest and
support vector machine, to categorize apps into benign and malware.
We find that random forest achieves the higher accuracy rate of
99.34\% for the entire Drebin app dataset and 99.82\% for the portion
of the Drebin dataset which contains malware apps from three most
common malware families only, in addition to all the benign apps.
These results indicate that our approach is better than other existing
approaches based on permissions only.

The main contributions of this paper are twofold:

\begin{itemize}

\item We created feature sets of Android permissions, removed
unnecessary permissions with zero impact and applied two permission
ranking algorithms Chi-Square Test and Fisher's Exact Test to reduce
the dimensionality of the dataset without loss of information content.

\item We employed three classification algorithms namely decision
tree, random forest and support vector machine to evaluate the
effectiveness of ranked permissions. The experimental results show
that random forest can achieve an accuracy rate which is better than
that of other reported approaches.

\end{itemize}

This paper is organized as follows. Section~\ref{sec:relatedwork}
introduces the related work. Section~\ref{sec:ourapproach} describes
our approach.  Experimental results are given in
Section~\ref{sec:experimentalresults} and compared with existing
methods in Section~\ref{sec:comparisonOthers}.
Section\ref{sec:conclusionandfuturework} concludes this paper.

\section{Related Work}
\label{sec:relatedwork}

Numerous authors have applied machine learning algorithms on Android
permissions to detect malware apps. Risks incurred by individual
permissions as well as by combinations of permissions has been
analyzed in \cite{wang2014exploring}, where feature ranking methods
such as mutual information, correlation coefficient and t-test were
employed to rank permissions based on their risk.  They also use
sequential forward selection and principal component analysis to
identify subsets of risky permissions and evaluate the effectiveness
of model using decision tree, random forest and support vector
machine.  Zhang \emph{et al.} \cite{zhang20:access} exploited function
call grah vectorization to extract information that allowed files to
be identified as malicious.

Drebin \cite{arp2014drebin} performs static analysis of Android apps
and identify malicious apps directly on smartphone. It gathers as many
features of an app from manifest file and disassembled code such as
hardware components, requested permissions, used permissions, app
components, filtered intents, restricted API calls, suspicious API
calls, network addresses. It embeds them in a joint vector space such
that typical patterns indicative for malware can be automatically
identified and used for explaining the decision. However, using large
number of features increases the complexity of the algorithm and hence
increases the computational overhead. Drebin performs binary
classification using a linear support vector machine that determines a
hyperplane and separates benign and malware apps with maximal margin. 

In our previous work \cite{saleem2020examining}, we have used kernel
density estimation, a non-parametric method for estimating probability
distribution from data, to examine Android permission patterns and
calculate the moments of permissions distribution. While examining
dangerous permissions, we found that benign apps have lower mean,
higher coefficient of variance, higher skewness and higher kurtosis
than malware apps.

Aung and Zaw \cite{zarni2013permission} proposed a framework to gather
permissions from manifest files of 500 Android APK files and develop a
machine learning-based malware detection system to detect Android
malware. Their system monitors permission-based features and events
extracted from Android apps and analyses these features by using
classification algorithms to distinguish between benign and malware
apps.

A number of Android apps, some 175 malware and 796 benign ones, have
been analyzed in \cite{chan2014static}. App permissions have been
extracted from manifest file and API calls from classes.dex file, and
stored in a feature set. They selected 19 relevant API calls using
information gain \cite{russellp} and compared the results using
machine learning classifiers.

Two main aspects of permission-based Android malware detection:
feature selection methods and classification algorithms have been
analyzed in \cite{pehlivan2014analysis}. They investigated four
different feature selection methods such as Gain Ratio Attribute
Evaluator, Relief Attribute Evaluator, Control Flow Subset Evaluator
and Consistency Subset Evaluator with machine learning classifiers.
Using these feature selection methods, they selected 97 features that
could represent the whole dataset.  Their experiments show that random
forest and J48 decision tree classifiers achieve the highest
performance of all feature selection methods used.

Forensic analysis tool FAMOUS (forensic analysis of mobile devices
using scoring of application permissions) was developed for the
detection of Android malware based on static analysis
\cite{kumar2018famous}.  This tool will scan all the installed apps of
Android device and provide an explanatory report. Their best
performing model uses weighted score feature set to achieve
classification based on app permissions.

An innovative malware detection framework for Android has looked into
more than 29,000 benign and malware apps in the period 2010 to 2019 to
identify the most significant permissions \cite{mathur21:jisa}.  The
authors evaluated eight ML algorithms and founds that the Random
Forest classifier based model performed best

We note that all of the different machine learning algorithms used did
not achieve high accuracy, with best results ranging from 91.75\% to
at most 94.84\% which is not too high \cite{zarni2013permission,
chan2014static,pehlivan2014analysis,kumar2018famous}.  In addition,
none of the above approaches can learn effective feature interaction,
which shows the improvement of the performance of learning models.
This observation has motivated us to find a better approach for
malware detection using app permissions.

\section{Our Approach: Relevant permission-based malware detection
system} \label{sec:ourapproach}

For our malware detection system, we used Drebin \cite{arp2014drebin}
dataset which contains data for a total of 114,298 benign apps and
5510 malware apps. Malware apps used in this dataset are classified
into 169 malware categories. The list of features used by each app
includes hardware components (GPS, Camera, Touchscreen etc.),
permissions, API calls, network addresses and others. 

From this dataset, we considered only app permissions and created two
datasets: one with all the apps and another with all benign apps and
malware apps from top three malware families only.  The malware in
this set include samples from FakeInstaller, DroidKungFu, and Opfake
families with 925, 666, and 612 malware samples, respectively, with a
total of 2203 samples or about 40\% of all malware apps.  As shown in
Table~\ref{experiments-dataset1}, our first dataset contains 119,808
apps and second dataset contains 116,501 apps.

\begin{table}[!h]
\caption{Datasets used in our experiments.}
\label{experiments-dataset1}
\centering
\begin{tabular}{ccccc} 
 \hline
dataset & total apps & benign apps & malware apps & description \\
\hline
Dataset1 & 119808 & 114298 & 5510 & All benign and all malware  \\
Dataset2 & 116501 & 114298 & 2203 & All benign and top
three malware families only \\
\hline
\end{tabular}
\end{table}

Our detection system proceeds in several steps. First, we remove
irrelevant permissions from the feature set. Then, we rank the
features using feature selection methods, namely Chi-Square test and
Fisher's Exact test. In the final step, we apply classification
algorithms, namely, Decision Tree, Support Vector Machine and Random
Forest to classify apps into  benign and malware ones.

\subsection{Remove irrelevant permissions}

An app can execute a specific operation only if it has required
permissions. We created a feature set that contains total of 94
permissions, out of which 37 are normal, 28 are dangerous, and 29 are
signature permissions. While analyzing Drebin dataset, we found that
45 of those permissions that were never used in any benign or malware
app in the dataset such as ADD\_VOICEMAIL, BIND\_CARRIER\_SERVICES,
BIND\_TEXT\_SERVICE etc. We removed these permissions and kept the
remaining 49 relevant ones.

\subsection{Feature Selection}

Our next step is to find permissions which have strong association
with class variable named AppCategory in dataset1 and dataset2.
AppCategory is a class variable that categorizes an app as benign or
malware. Individual permission is used to predict the AppCategory
based on their p-value and it leads to remove the unrelated
permissions.  Focusing on most relevant permissions aims to reduce the
dimensionality of the dataset without loss of accuracy, and also to
subsequently improve the training speed of the detection algorithms.
We used two feature ranking methods (Chi-Square test and Fisher's
Exact test) in our approach.

\begin{itemize}

\item Chi-Square test \cite{kenney1939mathematics} is a statistical
hypothesis test used to determine the correlation between individual
permissions and the AppCategory.  In Chi-Square test, we are ranking
permissions based on their p-values and rejecting the null hypothesis,
when the p-value of permission feature is less than or equal to 0.05.
The threshold value of 0.05 is usually set to reject the null
hypothesis. The null hypothesis is that individual permissions and the
AppCategory are independent and the correlation with a p-value lower
than the given threshold value is statistically significant. Smaller
p-value indicates stronger evidence against the null hypothesis, so
the null hypothesis can be rejected.

\item Fisher's Exact test \cite{liu2007computational} is another
statistical method used to determine the association between two
categorical variables. Just like Chi-Square test, Fisher's Exact test
evaluates the null hypothesis of independence. As in the previous
case, we reject the null hypothesis when the p-value of permission
features is less than or equal to 0.05.  It is worth noting that
Fisher's test is one of the exact tests because the significance of
deviation from the null hypothesis such as p-value can be calculated
exactly, rather than relying on an approximation
\cite{mcdonald2009handbook}.

\end{itemize}

Table~\ref{permissions-pvalues} shows the reduced set of permissions
(with their p-values) obtained by applying feature ranking algorithms
on dataset1 and dataset2.  For dataset1, Chi-Square test and Fisher's
Exact test identify the same number of permissions (47) whose p-values
less than or equal to 0.05; for dataset2, Chi-Square test has 39
permissions whose p-values less than or equal to 0.05 while Fisher's
Exact test has 2 additional permission (READ\_SYNC\_STATS and
REORDER\_TASKS) for a total of 41 permissions.

\subsection{Machine Learning based Classification Algorithms}

We employed three supervised machine learning-based classification
algorithms, namely, decision tree, random forest, and support vector
machine, to classify benign and malware apps.

\begin{itemize}

\item Decision Tree is used in prediction problems where the outcome
belongs to one out of a limited set of categories
\cite{quinlan1986induction}.  It is based on divide and conquer
strategy to build a suitable tree from a given learning set which
contains a set of labeled instances. In our case, nodes of the
decision tree are permissions and leaves are the established class
labels. Decision tress are built as a set of rules during learning
process. These rules are then used to predict the classes of test
observations. Decision trees predict by evaluating which class is the
most common among the training data within the partition rather than
taking average in each partition \cite{desiontree-classification}.

\item Support Vector Machine \cite{burges1998tutorial} is a supervised
machine learning algorithm used for binary classification. In SVM,
each observation is plotted as a point in $N$-dimensional space and
finds the hyper plane that separates the $N$-dimensional space into
two classes. Here $N$ represents the number of permissions or
features.  Given the labeled training data, SVM estimates an optimal
hyperplane which classifies new samples.

\item Random Forest \cite{breiman2001random} is an ensemble learning
method for classification that constructs a large number of individual
decision trees at the time of training or, in other words, makes a
forest of decision trees. It is an ensemble learning method because it
combines a number of decision trees into one predictive model where it
reduces the overfitting by averaging the result. Each decision tree
predicts the app category (benign or malware) and the app category
with the most votes becomes the final prediction of the algorithm.
Like other ensemble learning algorithms, random forest often surpasses
a single tree in terms of accuracy of class.

\end{itemize}

\section{Results and Evaluation}
\label{sec:experimentalresults}

\subsection{Experimental Setup}

All experiments were conducted on a PC with Intel Core i7-9700 running
at 4.7GHz with 32GB RAM memory, using RStudio version 1.25033 software
\cite{team2015rstudio}.

\newpage
\begin{table}[!h]
\caption{Permissions and their p-values after applying Chi-Square Test
and Fisher's Exact Test on dataset1 and dataset2.}
\label{permissions-pvalues} \centering
\begin{tabular} { |m{20em} | m{2.0cm} m{2.0cm} | m{2.0cm} m{2.0cm} |}
 \hline
 & \multicolumn{2}{l|}{p-values from Chi-Square test} &
 \multicolumn{2}{l|}{p-values from Fisher's Exact Test} \\
 & dataset1 & dataset2 &  dataset1 & dataset2  \\
\hline
ACCESS\_COARSE\_LOCATION & 1.68E-37 & N/A & 1.07E-35 & N/A \\ 				
ACCESS\_FINE\_LOCATION & 9.86E-26 & 4.90E-03 & 1.11E-24 & 4.36E-03  \\ 				
ACCESS\_LOCATION\_EXTRA\_COMMANDS & 0 & 5.42E-120 & 2.83E-201 & 2.22E-70 \\
ACCESS\_NETWORK\_STATE & 1.67E-57 & N/A & 6.11E-59 & N/A \\ 				
ACCESS\_WIFI\_STATE & 0 & 4.20E-176 & 0 & 2.49E-135 \\
BATTERY\_STATS & 1.58E-08 & N/A & 3.64E-07 & N/A \\ 				
BIND\_WALLPAPER & 5.47E-03 & N/A & 8.50E-03 & N/A \\ 				
BLUETOOTH & 1.06E-76 & N/A & 1.39E-50 & N/A \\ 				
BLUETOOTH\_ADMIN & 2.08E-84 & 4.44E-03 & 6.72E-53 & 1.25E-03 \\ 				
CALL\_PHONE & 2.90E-17 & 8.91E-26 & 4.96E-16 & 1.52E-33 \\ 				
CAMERA & 4.72E-20 & 3.49E-31 & 4.62E-23 & 3.98E-47  \\ 				
CHANGE\_NETWORK\_STATE & 5.79E-201 & 5.25E-108 & 1.82E-122 & 1.10E-60 \\ 				
CHANGE\_WIFI\_MULTICAST\_STATE & 6.05E-03 & 4.88E-02 & 1.29E-03 & 1.71E-02 \\ 				
CHANGE\_WIFI\_STATE & 0 & 0 & 0 & 0 \\
CLEAR\_APP\_CACHE & 3.36E-48 & N/A & 9.95E-29 & N/A \\ 				
DISABLE\_KEYGUARD & 1.90E-299 & 1.71E-04 & 7.87E-171 & 1.69E-05 \\
EXPAND\_STATUS\_BAR & 1.20E-46 & 1.14E-02 & 1.37E-28 & 1.11E-03 \\ 				
GET\_ACCOUNTS & 6.37E-89 & 3.57E-12 & 2.77E-66 & 1.95E-17 \\ 				
GET\_PACKAGE\_SIZE & 1.35E-58 & N/A & 1.97E-34 & N/A \\ 				
INSTALL\_PACKAGES & 0 & 0 & 0 & 0 \\
INTERNET & 1.68E-57 & 4.54E-25 & 2.74E-73 & 3.41E-32 \\ 				
KILL\_BACKGROUND\_PROCESSES & 1.95E-42 & 6.63E-05 & 6.60E-30 & 3.30E-07 \\ 				
MODIFY\_AUDIO\_SETTINGS & 7.97E-04 & 1.97E-05 & 4.55E-04 & 1.12E-06 \\ 			
PROCESS\_OUTGOING\_CALLS & 5.41E-115 & 6.26E-03 & 8.10E-70 & 2.45E-03 \\ 				
READ\_CALENDAR & 1.40E-11 & 2.81E-13 & 4.90E-14 & 1.12E-22  \\ 				
READ\_CONTACTS & 0 & 3.08E-21 & 1.94E-231 & 1.54E-18 \\
READ\_EXTERNAL\_STORAGE & 2.46E-115 & 0 & 4.58E-77 & 1.03E-144 \\ 		
READ\_LOGS & 7.31E-117 & 6.12E-06 & 1.81E-85 & 6.76E-07 \\ 				
READ\_PHONE\_STATE & 0 & 0 & 0 & 0  \\
READ\_SMS & 0 & 0 & 0 & 0 \\
READ\_SYNC\_SETTINGS & N/A & 1.77E-02 & N/A & 6.93E-03 \\ 				
READ\_SYNC\_STATS & 4.34E-03 & N/A & 4.13E-04 & 2.50E-02 \\ 				
RECEIVE\_BOOT\_COMPLETED & 0 & 0 & 0 & 0 \\
RECEIVE\_MMS & 3.98E-203 & N/A & 2.97E-97 & N/A \\ 				
RECEIVE\_SMS & 0 & 0 & 0 & 0 \\
RECEIVE\_WAP\_PUSH & 0 & 5.98E-152 & 2.89E-133 & 1.15E-45 \\ 
RECORD\_AUDIO & 6.63E-04 & 1.27E-06 & 4.14E-04 & 6.19E-08 \\ 				
REORDER\_TASKS & N/A & N/A & N/A & 3.64E-02 \\ 				
SEND\_SMS & 0 & 0 & 0 & 0 \\
SET\_WALLPAPER & 2.26E-23 & 2.67E-21 & 7.80E-21 & 3.57E-30 \\ 				
SET\_WALLPAPER\_HINTS & 1.62E-32 & 1.09E-02 & 2.66E-21 & 1.11E-03  \\ 				
SYSTEM\_ALERT\_WINDOW & 0 & 0 & 4.93E-191 & 1.50E-163 \\	
VIBRATE & 6.19E-52 & 1.20E-39 & 2.10E-48 & 9.55E-47 \\ 				
WAKE\_LOCK & 0 & 6.29E-77 & 0 & 1.61E-64 \\
WRITE\_CALENDAR & 5.95E-07 & 7.58E-13 & 7.26E-08 & 9.44E-23 \\ 				
WRITE\_CONTACTS & 1.70E-121 & 2.24E-17 & 1.25E-88 & 6.38E-29 \\ 				
WRITE\_EXTERNAL\_STORAGE & 0 & 1.91E-192 & 0 & 3.02E-189 \\ 
WRITE\_SETTINGS & 2.01E-128 & 6.99E-19 & 4.13E-97 & 6.01E-16 \\ 				
WRITE\_SYNC\_SETTINGS & 3.33E-02  & 2.03E-02 & 3.08E-02 & 9.51E-03 \\ 				
\hline
\end{tabular}
\end{table}

All experiments were conducted with four sets of features, shown in
Table~\ref{featuresetslist}.  The raw dataset that contains all
permissions (feature set 1) is used for reference. After the removal
of irrelevant permissions from the datasets we conducted experiments
on second feature set which contains remaining 49 permissions.
Finally, we apply feature ranking algorithms such as Chi-Square test
and Fisher's Exact test to create third and fourth feature set which
contains permissions whose p-value is less than or equal to 0.05.

\begin{table}[!t]
\caption{Feature sets used in our experiments.}
\label{featuresetslist}
\centering
\begin{tabular}{lll} 
 \hline
dataset & feature count (filtering) & description \\ 
\hline
dataset1, 2 & 94 (none) & all permissions  \\ 		
\hline
dataset1, 2 & 49 (relevant only) & only relevant (i.e., used) permissions  \\ 		
\hline
dataset1 & 47 (Chi-Square)  &  only permissions with $p \leq 0.05$\\
dataset2 & 39 (Chi-Square)  &  by Chi-Square test \\
\hline
dataset1 & 47 (Fisher's Exact) &  only permissions with $p \leq 0.05$ \\
dataset2 & 41 (Fisher's Exact) &  by Fisher's Exact test \\
\hline
\end{tabular}
\end{table}

\subsection{Evaluation Metrics}

In this section, we present evaluation metrics used in our
experiments. Since our goal is to identify malware apps, the basic
measures are as follows:

\begin{itemize}

\item True Positives (TP) are malware apps that are correctly
classified as malware.

\item True Negatives (TN) are benign apps that are correctly classified
as benign.

\item False positives (FP) are benign apps that are incorrectly
classified as malware.

\item False negatives (FN) are malware apps that are incorrectly
classified as benign.

\end{itemize} 

Furthermore, We are using the following derived evaluation metrics
namely accuracy rate, false positive rate, false negative rate, true
positive rate, true negative rate, precision and F-score for
calculating the effectiveness of chosen classifiers.

True positive rate (also referred to as recall) is the proportion of
malware apps correctly classified. Given the number of true positive
and false negative, true positive rate can be obtained as
\begin{equation}
TPR = \frac{TP}{TP + FN} \label{eqtpr}
\end{equation}

True negative rate is the proportion of benign apps correctly
classified; it can be obtained as
\begin{equation} 
TNR = \frac{TN}{TN + FP} \label{eqtnr}
\end{equation}

False positive rate is the proportion of benign apps that are
incorrectly classified as malware; it can be obtained as
\begin{equation} 
FPR = \frac{FP}{FP + TN} \label{eqfpr} 
\end{equation}

False negative rate is the rate of malware apps that are incorrectly
classified as benign apps. Given the number of false negative and true
positive, FNR can be obtained as
\begin{equation} FNR = \frac{FN}{FN + TP} \label{eqfnr} \end{equation}

Accuracy is the rate of correctly predicted malware samples and
benign samples out of all samples in the dataset and can be obtained
as
\begin{equation} 
ACC = \frac{TP + TN}{TP + TN + FP +FN} \label{eqaccuracyrate} 
\end{equation}

Precision is the accuracy of malware apps correctly classified; it is
also referred to as positive predictive rate. It can be obtained as
\begin{equation} 
Precision = \frac{TP}{TP + FP} \label{eqprecision}
\end{equation}

Finally, F-Score is a weighted average of recall and precision. Closer the
value of F-Score to 100\%, better the classification is performed. As
we have precision and recall values, F-score can be obtained as
\begin{equation}
\mbox{\emph{F-Score}} = \frac{(1 + \alpha^2) * (Precision *
Recall)}{(\alpha^2 * Precision + Recall)} \label{eqfscore} 
\end{equation}
where $\alpha$ indicates the relative weight assigned to precision as
opposed to recall. We are using $\alpha$=1, which indicates that equal
weight is assigned to both precision and recall.

\subsection{Results and Analysis}

In all experiments, we divided our dataset into training set and
testing set, with 70\% of the data used for training classifier and
the remaining 30\% used for testing. We also logged the processing
time of all the classifiers. Processing time includes CPU time (user
time, system time) and elapsed time that the classifiers spend to
predict the app category. CPU time is the combination of user time and
system time, where user time is the amount of time spent for
predicting the app category and system time is the amount of time
spent by the kernel or the operating system on behalf of predicting
the app category. Elapsed time is the wall clock time to predict the
app category. User time, system time and elapsed time are measured in
seconds.

\subsubsection{Decision Tree Classifier}

Decision tree experiments were conducted on two datasets with four
feature sets. If we compare the decision tree experiments in
Table~\ref{dt-accuracy}, experiments with Chi-Square and Fisher's
Exact test-filtered dataset1 have better F-Score values (68.77\%) for
dataset1 than the same dataset but with no filtering and with relevant
features only where F-scores are 62.23\% and 64.11\%, respectively.

\begin{table}[!h]
\caption{Experimental results for Decision Tree classifier.  All
results given in percent.}
\label{dt-accuracy}
\centering
\begin{tabular}{cccccccccc} 
 \hline
dataset & feature count (filtering) & ACC & FPR & FNR & TPR & TNR & Precision & F-Score 
 \\
\hline
\multirow{2}{*}{dataset1} & 94 (none) & 98.77 & 0.93 & 23.61 & 76.39 & 99.07 & 52.50 & 62.23  \\
         & 49 (relevant only) & 98.86 & 0.83 & 23.61 & 76.39 & 99.17 &
55.22 & 64.11  \\ \hline
\multirow{2}{*}{dataset2} & 94 (none) & 98.99 & 0.79 & 20.20 & 79.8 & 99.21 & 52.49 &
63.32  \\
       & 49 (relevant only) & 99.01 & 0.78 & 20.20 & 79.8 & 99.22 & 52.88 & 63.60  \\
\hline
\multirow{2}{*}{dataset1} & 47 (Chi-Square) & 99.08 & 0.62 & 23.61 & 76.39 & 99.38 & 62.53 & 68.77  \\
         & 47 (Fisher's Exact) & 99.08 & 0.62 & 23.61 & 76.39 & 99.38 & 62.53 & 68.77  \\
\hline
\multirow{2}{*}{dataset2} & 39 (Chi-Square) & 99.33 & 0.26 & 24.30 & 75.70 & 99.74 & 83.45 & 79.39  \\
         & 41 (Fisher's Exact) & 99.38 & 0.20 & 24.30 & 75.70 & 99.80 & 86.88 & 80.90  \\	
\hline
\end{tabular}
\end{table}

The same observation holds for dataset2 with F-Score values of 79.39\%
and 80.90\% for the two filtered feature sets in
Table~\ref{dt-accuracy}, as compared to F-Score values of 63.32\% and
63.60\% obtained for dataset2 with no filtering and with relevant
features only. Note that Chi-Square and Fisher's Exact feature sets
both lead to the same results in Table~\ref{dt-accuracy} for dataset1
due to the identical feature set, as per
Table~\ref{permissions-pvalues}. 

Regarding computation time shown in Table~\ref{dt-performance},
filtering using Fisher's Exact test and subsequent classification with
Decision Tree take the least processing time, but the difference is
not very large.

\begin{table}[!h]
\caption{Computation times (in seconds) for Decision Tree classifier.}
\centering
\begin{tabular}{ccccc} 					
 \hline
data set &  feature count (filtering) & user & system & elapsed \\ \hline
\multirow{2}{*}{dataset1} & 94 (none) & 1.27 & 0.08 & 1.35 \\ 					
         & 49 (relevant only)         & 0.69 & 0.01 & 0.70 \\ \hline
\multirow{2}{*}{dataset2} & 94 (none) & 1.38 & 0.10 & 1.47 \\	
         & 49 (relevant only)         & 0.80 & 0.01 & 0.81 \\ \hline
\multirow{2}{*}{dataset1} & 47 (Chi-Square)  & 0.74 & 0.00 & 0.74 \\
         & 47 (Fisher's Exact)        & 0.69 & 0.00 & 0.69 \\ \hline
\multirow{2}{*}{dataset2} & 39 (Chi-Square)  & 0.64 & 0.02 & 0.65 \\
         & 41 (Fisher's Exact)        & 0.58 & 0.03 & 0.61 \\ \hline
\end{tabular}
\label{dt-performance}
\end{table}

\subsubsection{Support Vector Machine classifier}

A total of 16 experiments were conducted for SVM: eight experiments
were conducted before tuning the parameters, which is done using the
default SVM parameters in R \cite{svm-documentation}; the other eight
were conducted after tuning the parameters in which case SVM
parameters cost and gamma have been set at the optimized level
obtained through the tune \cite{tune-documentation} function in R.
Gamma parameter defines how far the impact of a single training
example reaches, with small and large values of gamma corresponding to
a Gaussian function with large and small variance, respectively.  Cost
parameter controls the impact of each individual support vector and
controls the training errors. 

After tuning the SVM parameters, we observe an improvement in
evaluation metrics values, Table~\ref{SVM-accuracy} that
F-Score for dataset1 with 94 permissions has the value of 65.51\%
before tuning the parameters, but after tuning F-Score value
significantly increase to 88.42\%. Same is true with other SVM
experiments.

\begin{table}[!h]
\caption{Experimental results for Support Vector Machine classifier.
All results given in percent.}
\label{SVM-accuracy}
\centering
\begin{tabular}{cccccccccc} 
 \hline
dataset & feature count (filtering) & tuning & ACC & FPR & FNR & TPR & TNR & Precision & 	F-Score
 \\
\hline
\multirow{4}{*}{dataset1} & \multirow{2}{*}{94 (none)} & before  & 97.88 & 1.79 & 15.59 & 84.41 & 98.21 & 53.53 & 65.51   \\		
         &           & after   & 98.98 & 0.72 & 7.92 & 92.08 & 99.28 & 85.05
& 88.42  \\ \cline{2-10}
  & \multirow{2}{*}{49 (relevant only)} & before  & 98.19 & 1.53 & 12.48 & 87.52 & 98.47 & 60.49 & 71.53  \\	
         &           & after   & 99.13 & 0.56 & 7.93 & 92.07 & 99.44 & 87.99
& 89.99  \\ \hline
\multirow{4}{*}{dataset2} & \multirow{2}{*}{94 (none)} & before  & 99.25 & 0.55 & 13.41 & 86.59 & 99.45 & 71.54 & 78.35  \\ 					
         &           & after   & 99.62 & 0.28 & 5.79 & 94.21 & 99.72 & 86.85
& 90.38   \\ \cline{2-10}
  & \multirow{2}{*}{49 (relevant only)} & before  & 99.64 & 0.22 & 7.85 & 92.15 & 99.78 & 88.47 & 90.27  \\ 		
         &           & after   & 99.66 & 0.24 & 5.71 & 94.29 & 99.76 & 88.51 & 91.30  \\ 
\hline
\multirow{2}{*}{dataset1} & \multirow{2}{*}{47
(Chi-Square/Fisher's Exact)} & before & 98.49 & 1.29 & 7.35 & 92.65 & 98.71 & 72.96 & 81.63   \\
         &            & after  & 99.27 & 0.42 & 7.76 & 92.24 & 99.58 &
90.66 & 91.44   \\
\hline
\multirow{2}{*}{dataset2} & \multirow{2}{*}{39 (Chi-Square)
/ 41 (Fisher's Exact)} & before & 99.72 & 0.14 & 8.06 & 91.94 & 	99.86 & 92.51 & 92.22   \\ 
         &            & after  & 99.77 & 0.05 & 8.92 & 91.08 & 99.95 & 97.16 & 94.02  \\
\hline
\end{tabular}
\end{table}

We can also observe in Table~\ref{SVM-accuracy} that, after tuning the
parameters, SVM with Chi-Square and Fisher's Exact feature sets gives
the best results among SVM experiments.  In particular, it obtains the
best F-Score value of 91.44\% for dataset1, and 94.02\% for dataset2.
However, the computation time for SVM classifier, as shown in
Fig.~\ref{SVM-performance}, is much higher than that of its Decision
Tree counterpart.

\begin{table}[!h]
\caption{Computation time (in seconds) for Support Vector Machine
classifier.}
\label{SVM-performance}
\centering
\begin{tabular}{cccccc} 
 \hline
dataset & feature count (filtering) & tuning & user  & system & elapsed
 \\
\hline
\multirow{4}{*}{dataset1} & \multirow{2}{*}{94 (none)} & before & 40.45 & 0.03 & 40.48     \\ 
         &              & after  & 35.91 & 0.10 & 36.01     \\
	 \cline{2-6}
         & \multirow{2}{*}{49 (relevant only)} & before & 37.99 & 0 & 37.98     \\ 
         &       & after  &  32.59 & 0.03 & 32.63     \\ \hline
\multirow{4}{*}{dataset2} & \multirow{2}{*}{94 (none)} & before & 38.19 & 0.09 & 38.30     \\ 
         &  & after   & 33.76 & 0.02 & 33.78     \\  \cline{2-6}				
         & \multirow{2}{*}{49 (relevant only)} & before & 36.55 & 0.13 & 36.69     \\ 
         &  & after   & 31.09 & 0.08 & 31.15 \\ \hline					
\multirow{2}{*}{dataset1} & \multirow{2}{*}{47 (Chi-Square)} & before  & 35.42 & 0.12 & 35.55     \\
         &  & after   & 30.21 & 0.10 & 30.29     \\
	 \hline
\multirow{2}{*}{dataset2} & \multirow{2}{*}{39 (Chi-Square)} & before & 33.86 & 0.08 & 33.94     \\ 						
         &  & before  & 28.39 & 0.09 & 28.49     \\ \hline					
\multirow{2}{*}{dataset1} & \multirow{2}{*}{47 (Fisher's Exact)} & before & 32.99 & 0.17 &  33.17    \\ 						
         &  & after  & 29.08 & 0.14 & 29.24     \\
	 \hline
\multirow{2}{*}{dataset2} & \multirow{2}{*}{41 (Fisher's Exact)} & before  & 32.36 & 0.11 & 32.48     \\ 						
         &  & after  & 29.14 & 0.13 & 29.27     \\ \hline
\end{tabular}
\end{table}

\subsubsection{Random Forest Experiments}
\label{randomForestSection}

Random Forest experiments were conducted on two datasets with four
feature sets. As seen from Table~\ref{RF-performance}, Random Forest
with Chi-Square feature sets and Random Forest with Fisher's Exact
feature sets give the best results among all classifiers.  As can be
see, F-score values for dataset1 and dataset2 with Chi-Square
filtering are 92.03\% and 95.05\%, respectively, with the
corresponding accuracy rates of 99.35\% and 99.81\%, respectively.
When Fisher's Exact test is used, accuracy rate and F-score are
slightly higher at 99.34\% and 92.17\%, respectively, for dataset1,
and 99.34\% and 99.82\%, respectively, for dataset2.  While accuracy
values are virtually identical for both Chi-Square and Fisher's Exact
test filtering, the latter gives higher values for the F-score.
Accuracy is marginally better for dataset1 with Chi-Square filtering
on account of slightly better false positive rate, true negative rate
and precision rate.

As shown in Tables~\ref{dt-accuracy}, \ref{SVM-accuracy}, and
\ref{RF-accuracy}, the value of F-score increases with all classifiers
when we filter permissions, first by removing the irrelevant ones and
then by ranking the remaining ones and retaining those with p-value
below 0.05 using either Chi-Square or Fisher's Exact tests.  In fact,
the value of F-score reach to the maximum in all the classifiers.

\begin{table}[!t]
\caption{Experimental results for Random Forest classifier.
All results given in percent.}
\label{RF-accuracy}
\centering
\begin{tabular}{ccccccccc} 
 \hline
dataset & feature count (filtering) & ACC & FPR & FNR & TPR & TNR &
Precision & F-Score \\ \hline
\multirow{2}{*}{dataset1} &  94 (none) & 98.60 & 1.25 & 5.62 & 94.38
& 98.75 & 74.02 & 82.97  \\ 
                          & 49 (relevant only) & 98.89 & 0.88 & 6.64
			  & 93.36 & 99.12 & 81.71 & 87.15  \\ \hline
\multirow{2}{*}{dataset2} &  94 (none) & 99.60 & 0.33 & 4.45 & 95.55
& 99.67 & 82.79 & 88.71  \\ 
                          & 49 (relevant only) & 99.74 & 0.18 & 4.97 & 95.03 & 99.82 & 90.94 & 92.94  \\ 					
\hline
dataset1 & 47 (Chi-Square)  & 99.35 & 0.40 & 6.71 & 93.29 & 99.60 &
90.80 & 92.03  \\ \hline
dataset2 & 39 (Chi-Square)  & 99.81 & 0.03 & 8.30 & 91.70 & 99.97 &
98.65 & 95.05  \\ \hline
dataset1 & 47 (Fisher's Exact) & 99.34 & 0.56 & 3.16 & 96.84 & 99.44
& 87.93 & 92.17  \\ \hline
dataset2 & 41 (Fisher's Exact)  & 99.82 & 0.05 & 6.66 & 93.34 & 99.95
& 97.30 & 95.28  \\ \hline
\end{tabular}
\end{table}

\begin{table}[!t]
\caption{Computation time (in seconds) for Random Forest classifier.}
\label{RF-performance}
\centering
\begin{tabular}{cccccc}
 \hline
data set &  feature count (filtering) & user & system & elapsed \\ \hline
 \\ \hline
\multirow{2}{*}{dataset1} & 94 (none) & 2.44 & 0.06 & 2.50  \\ 					
                 & 49 (relevant only) & 2.30 & 0.03 & 2.33  \\ \hline						
\multirow{2}{*}{dataset2} & 94 (none) & 1.91 & 0.01 & 1.93  \\ 					
                 & 49 (relevant only) & 1.36 & 0.01 & 1.38  \\ \hline			
dataset1 & 47 (Chi-Square)            & 1.87 & 0.01 & 1.89  \\ \hline							
dataset2 & 39 (Chi-Square)            & 1.17 & 0.02 & 1.19  \\ \hline			
dataset1 & 47 (Fisher's Exact)        & 1.98 & 0.03 & 2.02  \\ \hline			
dataset2 & 41 (Fisher's Exact)        & 2.51 & 0.04 & 2.59  \\ \hline
\end{tabular}
\end{table}

In the above experiments, Decision Tree is giving the worst results in
terms of accuracy and F-score, while Random Forest is giving the best
results. If the malware is from the top three malware family then
Random Forest with Fisher's Exact feature set has the best F-Score of
95.28\% among all the classifiers. Even if the malware is not from the
top three malware family then still Random Forest with Fisher's Exact
feature set gives the best F-Score of 92.17\%. If we compare the
computational performance of all classifiers, then Decision Tree is
giving the best results and have processing time under 1.50 seconds,
but it is not good in accuracy and F-score. While Random Forest has
processing time under 2.60 seconds, which is also considered to be
good result because Random Forest operate by constructing a multitude
of decision trees and Random Forest also has the best accuracy rate
and F-score value among other classifiers.

\section{Comparison with existing methods}
\label{sec:comparisonOthers}

We evaluate our results by comparing them with the existing methods
and we find that some approaches do not work well in detecting Android
malware. There are many tools which depend on signatures [6] and look
for patterns and if specific pattern is not matched, then the
technique will not be able to detect that specific type of malware.
There are also some approaches which focus on risky permissions only
and ignore the normal permissions, and hence does not correctly
classify the benign and malware apps. There are few approaches who
apply their algorithms on small dataset and may lead to incorrect
classification results.

Our approach is more efficient than Drebin \cite{arp2014drebin} in
terms of accuracy because when we combine permissions with feature
ranking algorithms, we are able to detect malware with an accuracy
rate of 99.34\% for dataset1 and 99.82\% for dataset2 using Fisher's
Exact algorithm and random forest classifier while Drebin has the
accuracy rate of 93.90\% for full dataset and 95.90\% for Malgenome
dataset. 

Different ranking algorithms were used in \cite{wang2014exploring} but
they only use the high-risk permissions using sequential forward
selection and principal component analysis approaches, and ignore the
low risk permissions.  Contrary to that, we focus on all permissions,
regardless of whether they are categorized as normal permissions,
dangerous permissions or signature permissions, and retain only the
relevant ones. The best F-score value reported in
\cite{wang2014exploring} is about 91\% with the false positive rate of
0.60\% while our best F-score value is 92.17\% for dataset1 and
95.28\% for dataset2 with the false positive rate of 0.56\% and 0.05\%
respectively. Their target was to detect the malware abuse by
extracting high risk permissions, but our approach is to develop a
framework that can detect both benign apps and malware apps.

A combination of app permissions and source code-based analysis was
reported in \cite{milosevic2017machine}. They used different machine
learning classifiers such as C4.5 decision tree, random forest, SVM
etc. With the combination of classifiers, they achieved a best F-score
value of 95.6\% using source code-based analysis, which is slightly
better than our best F-score value 95.28\%. But they did the
experiments on small dataset, they only used 387 apps for
permissions-based analysis and 368 apps for source code-based
analysis. While our dataset contains a large number of apps, which
contains a total of 119,808 apps, 114,298 are benign apps and 5510 are
malware apps and the detection of Android malware from large number of
apps is a challenging task.

We also compare our results with APK Auditor \cite{talha2015apk},
which is a permission-based Android malware detection tool and
consists of three main parts, namely signature database, an Android
client, and central server which communicates between end user and
signature database. APK Auditor classifies apps based on permissions
and stores the extracted information of apps in a signature database.
However, they have a rather low accuracy rate of 88\% as opposed to
our accuracy rate of 99.34\% for dataset1 and 99.82\% for dataset2.
Their detection rate is low because they are using a total of 145
permissions including irrelevant permissions; our detection approach
is based on the most relevant features and is thus able to achieve a
higher accuracy rate.

The best results were achieved with Random Forest Classifier reported
in \cite{mathur21:jisa} which reported accuracy of 97\%, but that
approach does not involve feature interaction which is addressed in
our model.

Despite the accuracy rate higher than other approaches, we also
experience a false negative rate of 6.66\% for dataset2 and recall
rate of 93.34\%. In future, our aim is to improve these numbers with
the combination of other features such as API calls, method calls etc.

\section{Conclusion and Future Work}
\label{sec:conclusionandfuturework}

Permission system is one of the security procedures used by the
Android operating system. Our target was to develop an approach based
on the Android permission system for the detection of Android malware.
Our proposed approach utilized machine learning classifiers, which
were trained with the proposed features sets. Unnecessary feature sets
were removed from the datasets and then ranked using Chi-Square test
and Fisher's Exact test feature ranking algorithms. These feature sets
were ranked based on their p-values. Different experiments were
conducted on four feature sets and we concluded that the permissions
interactions based on Chi-Square and Fisher's Exact are the most
effective in detecting malware apps. If the malware app is from the
top three malware family, our approach produces the best results. Even
if the malware app is not from the top three malware family, still our
approach is better than many existing approaches. Comparison with the
state-of-the-art approaches are also done in this paper. Our
experimental results validate our approach for malware detection,
which can effectively detect malware with more accuracy and higher
F-Score compared to existing approaches. The experiment results show
that our approach combined with Fisher's Exact and random forest
algorithm has a high accuracy rate and F-score value. For future
research, our aim is to increase the precision rate and recall rate,
and hence increase F-score value with the combination of permissions
and other features such as API calls and methods calls, among others.

\section{Acknowledgments}\label{sec11}

Research presented here was in part supported through Canada's
National Science and Engineering Research Council (NSERC) Discovery
Grants.

\bibliographystyle{unsrt}

\begin{thebibliography}{10}

\bibitem{kelly2014report}
Gordon Kelly.
\newblock Report: 97\% of mobile malware is on android. this is the easy way
  you stay safe.
\newblock {\em Forbes Tech}, 2014.

\bibitem{gdata}
GData.
\newblock New malware every 10 seconds!
\newblock
  \url{https://www.gdatasoftware.com/blog/2018/05/30735-new-malware-every-10-seconds}.
\newblock last accessed January 20, 2022.

\bibitem{arp2014drebin}
Daniel Arp, Michael Spreitzenbarth, Malte H\"{u}bner, Hugo Gascon, and Konrad Rieck.
\newblock Drebin: Effective and explainable detection of android malware in
  your pocket.
\newblock In {\em Ndss}, volume~14, pages 23--26, 2014.

\bibitem{dash2016droidscribe}
Santanu~Kumar Dash, Guillermo Suarez-Tangil, Salahuddin Khan, Kimberly Tam,
  Mansour Ahmadi, Johannes Kinder, and Lorenzo Cavallaro.
\newblock Droidscribe: Classifying android malware based on runtime behavior.
\newblock In {\em 2016 IEEE Security and Privacy Workshops (SPW)}, pages
  252--261. IEEE, 2016.

\bibitem{yan2012droidscope}
Lok~Kwong Yan and Heng Yin.
\newblock Droidscope: Seamlessly reconstructing the $\{$OS$\}$ and dalvik
  semantic views for dynamic android malware analysis.
\newblock In {\em Presented as part of the 21st $\{$USENIX$\}$ Security
  Symposium ($\{$USENIX$\}$ Security 12)}, pages 569--584, 2012.

\bibitem{android-permissions}
Google.
\newblock Permissions overview.
\newblock
  \url{https://developer.android.com/guide/topics/permissions/overview}.
\newblock last accessed January~22, 2022.

\bibitem{wijesekera2015android}
Primal Wijesekera, Arjun Baokar, Ashkan Hosseini, Serge Egelman, David Wagner,
  and Konstantin Beznosov.
\newblock Android permissions remystified: A field study on contextual
  integrity.
\newblock In {\em 24th $\{$USENIX$\}$ Security Symposium ($\{$USENIX$\}$
  Security 15)}, pages 499--514, 2015.

\bibitem{wang2014exploring}
Wei Wang, Xing Wang, Dawei Feng, Jiqiang Liu, Zhen Han, and Xiangliang Zhang.
\newblock Exploring permission-induced risk in android applications for
  malicious application detection.
\newblock {\em IEEE Transactions on Information Forensics and Security},
  9(11):1869--1882, 2014.

\bibitem{zhang20:access}
Yipin Zhang, Xiaolin Chang, Yuzhou Lin, Jelena Mi\v{s}i\'{c}, and Vojislav~B.
  Mi\v{s}i\'{c}.
\newblock Exploring function call graph vectorization and file statistical
  features in malicious {PE} file classification.
\newblock {\em IEEE Access}, 8:44652--44660, March 2020.

\bibitem{saleem2020examining}
Muhammad~Suleman Saleem, Jelena Mi{\v{s}}i{\'c}, and Vojislav~B
  Mi{\v{s}}i{\'c}.
\newblock Examining permission patterns in android apps using kernel density
  estimation.
\newblock In {\em 2020 International Conference on Computing, Networking and
  Communications (ICNC)}, pages 719--724. IEEE, 2020.

\bibitem{zarni2013permission}
Zarni Aung and Win Zaw.
\newblock Permission-based android malware detection.
\newblock {\em International Journal of Scientific \& Technology Research},
  2(3):228--234, 2013.

\bibitem{chan2014static}
Patrick~PK Chan and Wen-Kai Song.
\newblock Static detection of android malware by using permissions and api
  calls.
\newblock In {\em 2014 International Conference on Machine Learning and
  Cybernetics}, volume~1, pages 82--87. IEEE, 2014.

\bibitem{russellp}
Stuart Russell and Peter Norvig.
\newblock {\em Artificial Intelligence: A Modern Approach}.
\newblock Prentice Hall, third edition, 2010.

\bibitem{pehlivan2014analysis}
U{\u{g}}ur Pehlivan, Nuray Baltaci, Cengiz Acart{\"u}rk, and Nazife Baykal.
\newblock The analysis of feature selection methods and classification
  algorithms in permission based android malware detection.
\newblock In {\em 2014 IEEE Symposium on Computational Intelligence in Cyber
  Security (CICS)}, pages 1--8. IEEE, 2014.

\bibitem{kumar2018famous}
Ajit Kumar, KS~Kuppusamy, and G~Aghila.
\newblock Famous: Forensic analysis of mobile devices using scoring of
  application permissions.
\newblock {\em Future Generation Computer Systems}, 83:158--172, 2018.

\bibitem{mathur21:jisa}
Akshay Mathur, Laxmi~Mounika Podila, Keyur Kulkarni, Quamar Niyaz, and Ahmad~Y
  Javaid.
\newblock {NATICUSdroid}: A malware detection framework for android using
  native and custom permissions.
\newblock {\em Journal of Information Security and Applications}, 58:102696,
  2021.

\bibitem{kenney1939mathematics}
John~F Kenney.
\newblock {\em Mathematics of statistics}.
\newblock D. Van Nostrand, 1939.

\bibitem{liu2007computational}
Huan Liu and Hiroshi Motoda.
\newblock {\em Computational methods of feature selection}.
\newblock CRC Press, 2007.

\bibitem{mcdonald2009handbook}
John~H McDonald.
\newblock {\em Handbook of biological statistics}, volume~2.
\newblock sparky house publishing Baltimore, MD, 2009.

\bibitem{quinlan1986induction}
J~Ross Quinlan.
\newblock Induction of decision trees.
\newblock {\em Machine learning}, 1(1):81--106, 1986.

\bibitem{desiontree-classification}
Reza Hashemi.
\newblock Classification (decision) trees.
\newblock
  \url{http://rstudio-pubs-static.s3.amazonaws.com/521300_d4eea870fd224bc5bfe0cc1f1bc7b468.html}.
\newblock last accessed January 20, 2022.

\bibitem{burges1998tutorial}
Christopher~JC Burges.
\newblock A tutorial on support vector machines for pattern recognition.
\newblock {\em Data mining and knowledge discovery}, 2(2):121--167, 1998.

\bibitem{breiman2001random}
Leo Breiman.
\newblock Random forests.
\newblock {\em Machine learning}, 45(1):5--32, 2001.

\bibitem{team2015rstudio}
RStudio Team et~al.
\newblock Rstudio: integrated development for r.
\newblock {\em RStudio, Inc., Boston, MA URL http://www. rstudio. com}, 42:14,
  2015.

\bibitem{svm-documentation}
Rdocumentation.
\newblock Support vector machines.
\newblock
  \url{https://www.rdocumentation.org/packages/e1071/versions/1.7-3/topics/svm}.
\newblock last accessed January 20, 2022.

\bibitem{tune-documentation}
Rdocumentation.
\newblock Parameter tuning of functions using grid search.
\newblock
  \url{https://www.rdocumentation.org/packages/e1071/versions/1.7-3/topics/tune}.
\newblock last accessed January 20, 2022.

\bibitem{milosevic2017machine}
Nikola Milosevic, Ali Dehghantanha, and Kim-Kwang~Raymond Choo.
\newblock Machine learning aided android malware classification.
\newblock {\em Computers \& Electrical Engineering}, 61:266--274, 2017.

\bibitem{talha2015apk}
Abdullah Talha Kabakus, Ibrahim Alper Dogru, and Cetin Aydin.
\newblock APK auditor: Permission-based android malware detection system.
\newblock {\em Digital Investigation}, 13:1--14, 2015.

\end{thebibliography}

\end{document}